# Tunable Chirality-dependent Nonlinear Electrical Responses in 2D Tellurium

Chang Niu[1,2]†, Gang Qiu[1,2]†, Yixiu Wang[3], Pukun Tan[1,2], Mingyi Wang[3], Jie Jian[4], Haiyan Wang[4], Wenzhuo Wu[3] and Peide D. Ye[1,2]*

[1]*Elmore Family School of Electrical and Computer Engineering, Purdue University, West Lafayette, IN 47907, United States.*

[2]*Birck Nanotechnology Center, Purdue University, West Lafayette, IN 47907, United States.*

[3]*School of Industrial Engineering, Purdue University, West Lafayette, IN 47907, United States.*

[4]*School of Materials Science and Engineering, Purdue University, West Lafayette, Indiana 47907, United States.*

†These authors contributed equally to this work: Chang Niu, Gang Qiu.

*Correspondence and requests for materials should be addressed to P. D. Y. (yep@purdue.edu)

**Abstract:**

Tellurium (Te) is an elemental semiconductor with a simple chiral crystal structure. Te in a two-dimensional (2D) form synthesized by solution-based method shows excellent electrical, optical, and thermal properties. In this work, the chirality of hydrothermally grown 2D Te is identified and analyzed by hot sulfuric acid etching and high-angle tilted high-resolution scanning transmission electron microscopy. The gate-tunable nonlinear electrical responses, including the nonreciprocal electrical transport in the longitudinal direction and the nonlinear planar Hall effect in the transverse direction, are observed in 2D Te under a magnetic field. Moreover, the nonlinear electrical responses have opposite signs in left- and right-handed 2D Te due to the opposite spin polarizations ensured by the chiral symmetry. The fundamental relationship between the spin-orbit coupling and the crystal symmetry in two enantiomers provides a viable platform for realizing chirality-based electronic devices by introducing the chirality degree of freedom into electron transport.

**Keywords: Chirality, Nonlinear transport, 2D Tellurium, Spin Texture, Gate-tunability**

Chirality, a fundamental geometric property, emerges as an important topic in many areas, including physics, chemistry, and biology. In condensed matter physics, the chiral crystal structure couples with the electrons[1–4] by introducing a new chirality degree of freedom into the transport. Nonlinear electrical effects[5,6] induced by the broken inversion symmetry are suitable for detecting novel chirality-dependent phenomena in non-magnetic strong spin-orbit coupling materials with two opposite chiral crystal structures related by mirror symmetry, providing an opportunity for realizing chirality-based electronic devices.

Tellurium (Te) is an elemental narrow band gap semiconductor with a simple chiral crystal structure. Covalently bonded Te atomic chains form a trigonal crystal lattice through van der Waals interaction, as shown in **Figure 1a**. Te has been known to have a chiral crystal structure since 1924.[7] The two enantiomers related by mirror symmetry fall into two different space groups P3$_1$21 (right-handed) and P3$_2$21 (left-handed). The highest occupied state and the lowest unoccupied state are located around the corners of the Brillouin zone *H* and *H'* point (**Figure 1b**). The strong spin-orbital coupling (SOC)[8] splits the conduction band and forms a Weyl node at *H* (*H'*) point.[9–11] Due to the low symmetry of the chiral crystal structure and the Weyl-type SOC,[4] Te has a nontrivial radial spin texture where the spin polarization is parallel to the $k$ direction in the conduction band.[9,12–14] The radial spin texture is different from that in topological insulator surface states and Rashba semiconductors, as shown in **Figure S1**. In the valence band,[15] the camelback structure also inherits a spin texture with spin polarization nearly parallel to the $k_z$ direction (helical chain direction). The energy dispersion of the two enantiomers is the same as shown in **Figure 1c**. However, because of the mirror symmetry, the spin polarization (indicated by the arrows)[16] and the monopole charge of the Weyl nodes are opposite.[3] The chirality-dependent nonlinear electrical responses will arise when applying an external magnetic field which breaks the time-reversal symmetry and serves as a perturbation to the spin-splitting bands. Two-dimensional (2D) Te grown by the hydrothermal method[17] has attracted lots of attention recently due to its attractive electrical,[18] thermal,[19] and optical[20] properties. However, the chirality-related properties of hydrothermally grown 2D Te remain unexplored.

In this paper, we report the first systematic study of the gate-controlled chirality-dependent second-order nonlinear electrical responses under a magnetic field originating from the broken chiral symmetry of the crystal structure, including the nonreciprocal electrical transport (longitudinal direction) and the nonlinear planar Hall effect (transverse direction) in 2D Te. The chirality of the hydrothermally grown 2D Te is also studied by hot sulfuric acid etching and further confirmed by high-resolution scanning transmission electron microscopy (HR-STEM) images taken under the high-angle annular dark field (HAADF) mode and high-angle tilt condition. Our results provide a coherent approach to study the chirality-induced physical properties in 2D Te and other chiral materials.

The device structure of 2D Te field-effect transistors (FETs) used for nonlinear electrical measurement is shown in **Figure 1d**. The carrier concentration can be electrostatically tuned using 90 nm $SiO_2$ as the back gate. An ac excitation $I$ is applied in the Te atomic chain direction with a frequency of $\omega$. The first-order ($\omega$) and second-order ($2\omega$) longitudinal ($V_{zz}$) and transverse ($V_{zx}$) voltage drops are measured.

**Results and Discussion**

**Gate-tunable nonreciprocal electrical transport in 2D Te.**

Nonreciprocal phenomena[21–27] describe the directional transport of particles, for example, electrons in p-n junctions and photons in chiral materials (naturally optical activity).[28] The electrical magnetochiral anisotropy (eMChA) [29] is a nonlinear nonreciprocal electrical response in non-centrosymmetric material systems. Chiral crystals in a magnetic field should exhibit nonreciprocal transport,[30] in which the currents moving in $+k$ and $-k$ direction are different because of the broken inversion and time-reversal symmetry. This unidirectional magnetoresistance is described by[31]

$$R^{l,r}(B,I) = R_0(1 + \beta B^2 + \gamma^{l,r} BI) \qquad (1)$$

where the second term $\beta B^2$ represents the usual magnetoresistance, the third term $\gamma^{l,r} BI$ is the nonreciprocal contribution, and $\gamma^{l,r}$ describes the strength of the eMChA in a left- and right-handed material. The tensor nature of eMChA is revealed in Te bulk material[32] due to the low symmetry of the crystal structure and the coupling between helical Te chains. It remains in 2D Te films also. Because of the 2D nature of the Te flakes, the carrier density and carrier type are tunable by applying a gate voltage. The nonreciprocal transport caused by the Weyl-type SOC-induced radial spin texture in the conduction band differs from that reported in the p-type bulk Te.[32]

We measured the four-terminal magnetoresistance of 2D Te along $z$ direction by applying the magnetic field in the $y$ direction (**Figure 2a**). The phase-sensitive measurement was used to obtain the resistance difference ($\Delta R \equiv R(B,I) - R(B,-I)$) between currents moving in $+I$ and $-I$ direction. An ac excitation current $I^\omega$ was injected into the device from the drain electrode. The voltage difference $V_{zz}^\omega$ between A and B electrodes and its second harmonic $V_{zz}^{2\omega}$ were measured as illustrated in **Figure 2a**. The normalized resistance difference $\frac{\Delta R}{R}$ which is linear in magnetic field $B$ and current $I$ can be calculated using

$$\frac{\Delta R}{R} = 2\gamma^{l,r} BI = \frac{4V_{zz}^{2\omega}}{V_{zz}^\omega} \qquad (2)$$

Because the nonreciprocal transport is odd in $B$, we took the difference between the results for $+B$ and $-B$ to eliminate the influence of other effects, for example, the asymmetry of the device structure.

The nonreciprocal electrical transport in a 2D Te FET is measured at the temperature of 100 mK. The magnetic field dependence of $\frac{\Delta R}{R}$ at different currents is shown in **Figure 2a**. **Figure 2b** shows the slope $\frac{\Delta R}{RB}$ (black square data points) calculated from the linear fitting of the data in **Figure 2a**. The good linear relationship between $\frac{\Delta R}{R}$ and $B$ ($I$) consists with the eMChA described by equation (2). The density-dependent eMChA of 2D Te is measured by applying the back gate voltage. Two different measurement configurations (**Figure 2c** inset) are used to characterize the unidirectional magnetoresistance. When the current is reversed, the opposite value of $\frac{\Delta R}{R}$ further confirms the observation of the eMChA in 2D Te. The slope $\frac{\Delta R}{RB}$ decreases with the increasing gate voltage as shown in **Figure 2d**. Similar current, magnetic field, and gate voltage dependence of $\frac{\Delta R}{R}$ were observed in p-type 2D Te flakes shown in **Figure S2 and S3**. Our measurements indicate that the nonreciprocal transport is large when the carrier density is relatively low (Fermi level around band edge). The eMChA is first demonstrated in the Bismuth helix,[29] a macrocosm experiment in one dimension. Te has the simplest chiral structure (three atoms in one period) on the atomic scale; as a result, the eMChA is expanded to three-dimension space when the external magnetic field couples the electron bands. The gate-tunable eMChA in 2D Te originates from the strong three-dimensional (3D) spin-orbit coupling in the inversion asymmetric Te crystal.[10] The radial spin texture is strong at the band edges.[9] The nonreciprocal response attenuates quickly after the Fermi level is tuned away from the camelback structure or the Weyl nodes, indicating that the spin polarization configuration of the band plays an important role in the nonreciprocal transport.[33] The angular-dependent nonreciprocal electrical transport in n-type 2D Te is measured in two different planes ($x$-$y$ plane and $z$-$y$ plane, where the $x$-axis is the two-fold rotation axis, and the $z$-axis is the three-fold screw axis) shown in **Figure 2e**. The data points are fitted well with cosine function, indicating the nonreciprocal transport in 2D Te is large when the magnetic field is along the $y$-direction, which is asymmetry. We observed a small ($2 \times 10^{-6}\ T^{-1}$) and near zero nonreciprocal current when the magnetic field is applied along $x$- and $z$-direction, respectively, as shown in **Figure S4**. The band evolution under a magnetic field is sensitive to the crystal orientation, implying the crystal symmetry enforced by 3D spin-orbit coupling bands in Te. It is not easy to realize in other materials beyond 2D Te by directly comparing the topological trivial band (valence band) and the topological non-trivial band (conduction band) in the same sample with tunable gating and Fermi level. We noticed that the amplitude of the nonreciprocal electrical transport in n-type 2D Te is about one or two orders of magnitude larger than the amplitude in p-type 2D Te. The enhancement of nonreciprocal transport in n-type Te might be related to the topological nature of the Te conduction band and the band evolution under the magnetic field. The conduction band of Te has large Berry curvature because of the Weyl node. The valence band has no band crossing or Weyl node near valence band maximum. The theoretical studies of the relationship between chiral anomaly and nonreciprocal transport were reported.[34]

**Nonlinear planar Hall effect in 2D Te.**

Besides the nonreciprocal electrical response in the longitudinal direction, we also observed the nonlinear planar Hall effect [35,36] in 2D Te by measuring the Hall signal ($V_{zx}$) between C and B electrodes under an in-plane ($x$-$z$ plane) magnetic field, as shown in **Figure 3a**. An ac excitation current $I^\omega$ was injected into the device along the atomic chain direction. The second-order Hall signal $V_{zx}^{2\omega}$ depending on the magnetic field angle $\alpha$ with the longitudinal current is generated. It follows a cosine angular dependence and has a period of 360° at a fixed magnetic field $B$ and current $I$. The $|V_{zx}^{2\omega}|$ has the largest value when the magnetic field and current direction are orthogonal ($\alpha = \pm 90°$) and becomes zero when the magnetic field and current direction are parallel to each other ($\alpha = 0°\ or\ -180°$). The nonlinear planar Hall effect observed in 2D Te differs from that in topological insulators,[35] which reaches the maximum when the magnetic field is antiparallel to the current direction and vanishes when the magnetic field is orthogonal to the current direction. Because in 2D Te, the spin-polarization is parallel to the electron momentum direction (**Figure 1c**). In contrast, the spin-polarization at the surface of the topological insulators is perpendicular to the electron momentum direction. The nonlinear planar Hall signal also fits the two-fold rotation ($C_2^x$) symmetry requirement,[36] which is $V_{zx}^{2\omega}(\alpha = 0°) = V_{zx}^{2\omega}(\alpha = 180°) = 0$. $V_{zx}^{2\omega}$ complies with $V_{zx}^{2\omega}(+B) = -V_{zx}^{2\omega}(-B)$ in the different polarization of a magnetic field, as shown in **Figure S5**. In order to exclude other artificial components, the second-harmonic planar Hall data is obtained by taking the average between positive and negative magnetic fields ($V_{zx}^{2\omega}(B) = 1/2(V_{zx}^{2\omega}(+B) - V_{zx}^{2\omega}(-B))$). The amplitude of the nonlinear planar Hall effect $\Delta V_{yx}^{2\omega}$ is extracted from the cosine fitting using $V_{zx}^{2\omega} = \Delta V_{zx}^{2\omega} \cos(\alpha)$. The potential capacitance coupling effect is excluded by changing the frequency of the input ac excitation (**Figure S6**), which shows no significant change from 7 Hz to 87 Hz.

The external magnetic field perturbs the Te conduction band and breaks the time-reversal symmetry, which is essential in generating the second-order Hall signal. With the increasing strength of the magnetic field $B$, $V_{zx}^{2\omega}$ also increased (**Figure 3a**). The extracted amplitude of the second-harmonic planar Hall signal $\Delta V_{zx}^{2\omega}$ is plotted as a function of the magnetic field $B$, as shown in **Figure 3b**, indicating the linear dependence between $V_{zx}^{2\omega}$ and $B$. **Figure 3c** shows the current-dependent nonlinear planar Hall effect. Solid lines are curves fitting with $V_{zx}^{2\omega} = \Delta V_{zx}^{2\omega} \cos(\alpha)$. The extracted normalized second-order Hall signal $\Delta V_{zx}^{2\omega}/I$ is linearly dependent on the current $I$ (**Figure 3d**), indicating the nonlinear characteristic of the planar Hall effect, while the first-order ordinary Hall effect ($\Delta V_{zx}^{\omega}/I$) is constant. The magnetic field and current dependence of the nonlinear planar Hall effect are tunable and consistent at different carrier densities (back gate voltages $V_{bg}$) shown in **Figure 3b, 3d, and S7**. According to the analysis of the experimental data above, the nonlinear planar Hall effect follows:

$$V_{zx}^{2\omega\ l,r}(B,I) = \chi^{l,r} B I^2 \qquad (3)$$

where $\chi^{l,r}$ describes the strength of the nonlinear planar Hall effect, which can be calculated from the linear fit in **Figure 3b** and **3d**. The effect is explained by the asymmetry spin-splitting Te conduction band along $k_x$ direction under the magnetic field, as illustrated in **Figure 3e**. When the magnetic field is turned on, the conduction band will distort due to the interaction between spin-polarized bands and the external magnetic field.[31] Therefore, a strong carrier density dependence of $\chi^{l,r}$ is expected in 2D Te. The gate-dependent $\chi^r$ is shown in **Figure 3e**. Three different regions are observed. The effect is strong at low carrier density (region I) due to the large asymmetry of the spin-splitting bands. With the increasing electron density, the nonlinear planar Hall effect changes sign (region II) because of the different strengths and the opposite contribution to the nonlinear planar Hall effect of the Weyl Fermions in the inner Fermi surface. The film becomes conducting at high carrier density (region III), and $|\chi^r|$ decreases. The density-dependent nonlinear planar Hall effect is the total response of the spin-orbit interaction and the Weyl Fermions. The top gate is used for tuning the electron density by setting the bake gate voltage at a relatively high value ($V_{bg} = 33\ V$) to ensure the ohmic contact. Similar back gate voltage dependence of the nonlinear planar Hall effect is shown in **Figure S8**.

**The chirality of hydrothermally grown 2D Te.**

As shown in **Figure 4a** and **4b** the chirality of the covalently bonded Te atomic chains determine the chirality of the 2D Te crystal. Left- or right-handed Te can be determined using the shape of asymmetric etch pits produced by hot sulfuric acid etching on the cleaved $(10\bar{1}0)$ surface of bulk Tellurium.[37–39] The chirality of Te not only affects the etching process but also affects the radial spin texture of electrons.[12,40-43] The opposite outward and inward radial spin textures are expected in left- and right-handed Te crystals.

Here, the chirality of hydrothermally grown 2D Te flakes is investigated using hot sulfuric acid etching (see methods for details). Well-defined asymmetry etch pits are found on the etched 2D Te flakes with two different shapes that are mirror images of each other shown in **Figure 4c** and **4d**, indicating the presence of both left- and right-handed 2D Te crystals. The dimension of the etch pits in 2D Te is usually 2-8 µm, which can be easily identified by the scanning electron microscope (SEM) and the standard optical microscope (**Figure S9-11**). The angles ($\alpha = 36.9°, \beta = 90.0°, \gamma = 126.9°$ and $\theta = 106.2°$) can be calculated from the lattice constants (**Figure S12**). The $z$ direction (helical chain direction) of 2D Te can be known through the direction of the longest side of the etch pits, in agreement with the previous report. The observation of the etch pits in hydrothermally grown 2D Te provides an ideal platform for the study of asymmetric chiral growth under different initial chiral imbalance conditions and growth environments.[44]

Different chirality is also found in 2D Te using HR-STEM imaging under HAADF mode (also called Z-contrast) in addition to etched pits. In order to determine the chirality, at least two sets of STEM images from different zone axes are required to holographically reconstruct the atom arrangement in 3D space. STEM images were first taken with incident electron beam along out-of-plane direction ($[10\bar{1}0]$ axis) to ensure the crystal orientation (configuration is shown in **Figure 4e**). The STEM images of two crystals with different chirality (**Figure 4f** and **4g**) show identical top views which can be perfectly replicated by the projection of three-fold screw helices (blue balls, see **Supporting Video 1**). A second set of STEM images was taken along $[11\bar{2}0]$ direction by rotating the samples along [0001] axis for 30º counterclockwise, as shown in **Figure 4h**. It is noted that now the STEM images along $[11\bar{2}0]$ axis of two crystals (**Figure 4i** and **4j**) are distinctive and linked by mirror symmetry. Furthermore, 3D atomic model is used to simulate the 30º rotational transformation of Te crystals for both left- and right-handedness, as shown in **Figure 4i** and **4j** (also in **Supporting Video 2** and **3**). Hence, by analyzing two sets of STEM images along different zone axes, the opposite chirality was confirmed. By the combination of high angle tilted HR-STEM and etched pits, the chirality of the 2D Te films can be conclusively determined.

**Chirality-dependent nonlinear electrical responses in 2D Te.**

In this session, we demonstrated the connection between the chiral crystal structure and the nonlinear electrical responses in 2D Te. As shown in **Figure 1c**, the $D_3$ symmetry of Te requires no net spin polarization in $k$ space. With the introduction of a magnetic field, the time-reversal symmetry is additionally broken, leading to the formation of asymmetric electronic bands ($E_\uparrow(+k) \neq E_\downarrow(-k)$) around the Fermi level, resulting in nonlinear electrical responses, where the spin current is partially converted into a charge current. The eMChA and the nonlinear planar Hall effect will both have opposite signs in left- and right-handed materials ($\gamma^l = -\gamma^r$ and $\chi^l = -\chi^r$), required by the parity reversal symmetry. The asymmetric band evolution under a magnetic field is also responsible for the gate-tunable nonlinear transport in 2D Te.

**Figure 5a** and **5b** show the nonreciprocal electrical transport in the left- and right-handed 2D Te, respectively. The unidirectional magnetoresistance is characterized by the linear dependence of $\frac{\Delta R}{R}$ with the increasing magnetic field. When the current flows in $+I$ ($-I$) direction, the opposite sign of $\frac{\Delta R}{R}$ in left- and right-handed 2D Te indicates the opposite spin-orbit coupling induced by the chiral crystal structure. 5 different devices are measured including 3 right-handed 2D Te and 2 left-handed 2D Te, showing the same electrical measurement results (**Figure S13 and S14**).

The chirality-dependent nonlinear planar Hall effect is also observed in 2D Te. **Figure 5c** and **5d** show the carrier density dependence of $\chi^{l,r}$ in the left- and right-handed 2D Te. The insets show the second-order Hall voltage $V_{zx}^{2\omega}$ as a function of the angle $\alpha$. The planar Hall effect is measured under the same experiment setup in different Te flakes.

**Figure S15** lists all four possible magnetic field and chirality configurations. The magnetic field *B* direction is always parallel (antiparallel) to the second-order Hall voltage direction in right- and left-handed 2D Te at low carrier density (region I). Three different devices were measured, including two right-handed 2D Te flakes and one left-handed 2D Te flake.

It is worth emphasizing that in Te, the electron spin is polarized along the current direction, which is different from that in Rashba spin-orbit coupling system[45] and 2D transition metal dichalcogenides,[46] providing another type of spin-charge configuration for the possible application in spintronic devices. The opposite gate-tunable second-order nonlinear responses between left- and right-handed 2D Te introduce another chirality degree of freedom in to the electron transport, making 2D Te a suitable candidate for chirality-based frequency doubling and rectification nonlinear electronic devices.[47]

In this paper, we identified the chirality of the hydrothermally grown 2D Te using the asymmetric etch pits, and high-angle tilted HR-STEM imaging technique. The observed gate-tunable chirality-dependent nonlinear electrical responses, including nonreciprocal electrical transport and nonlinear planar Hall effect in 2D Te, prove the fundamental relationship between the spin-orbit coupling and chiral crystal symmetry. Our work provides a new route using magneto-transport to investigate the chirality-dependent physical and topological properties in 2D Te and other chiral materials.

**Materials and Methods**

**Hydrothermal growth of 2D Te flakes.** 0.09 g of $Na_2TeO_3$ (Sigma-Aldrich) and 0.5 g of polyvinylpyrrolidone (PVP) (Sigma-Aldrich) were dissolved in 33 ml double-distilled water. 3.33 ml of aqueous ammonia solution (25-28%, w/w%) and 1.67 ml of hydrazine hydrate (80%, w/w%) were added to the solution under magnetic stirring to form a homogeneous solution. The mixture was sealed in a 50 ml Teflon-lined stainless steel autoclave and heated at 180 °C for 30 hours before naturally cooling down to room temperature.

**Sulfuric acid etching of 2D Te flakes.**

The synthesized 2D Te flakes were transferred onto a 90 nm $SiO_2$/Si substrate to ensure the etching direction. 2D Te flakes were cleaned following a DI water rinse and standard solvent cleaning process (acetone, methanol, and isopropanol). 2D Te flakes were etched in hot concentrated sulfuric acid at 100 °C for 5 min. The scanning electron microscope (SEM) images were taken by a Thermo Scientific Apreo S SEM system at 5 kV.

**High resolution scanning transmission electron microscopy (HR-STEM)**

TEM, selected area diffraction, and HAADF-STEM analysis were performed with FEI TALOS F200x. This microscope was operated with an acceleration voltage of 200 kV. A high angle tilt holder was used to allow the large angle tilting required for this work.

**Device fabrication.** Te flakes were transferred onto a 90 nm $SiO_2$/Si substrate. The four-terminal and Hall-bar devices were patterned using electron beam lithography, and metal contacts were deposited by electron beam evaporation. 20/60 nm Ti/Au and 20/60 nm Ni/Au were used as an electrical contact for p-type and n-type 2D Te contact, respectively. 20 nm ALD $Al_2O_3$ grown at 200 °C using $(CH_3)_3Al$ (TMA) and $H_2O$ as precursors were used to dope the 2D Te from p-type to n-type.

**Low temperature magneto-transport measurements.** The magneto-transport measurements were performed in a Triton 300 (Oxford Instruments) dilution fridge system with 12 T superconducting coils at a temperature down to 50 mK. A portion of the measurement which requires magnetic field rotation was performed in an 18 T superconducting magnet system (SCM2) in the National High Magnetic Field Laboratory in Tallahassee, Florida. The electrical data were acquired by standard small signal AC measurement technique using SR830 and SR860 lock-in amplifiers (Stanford Research).

**Acknowledgements**

P.D.Y. was supported by Army Research Office under grant No. W911NF-15-1-0574. W.W. acknowledges the School of Industrial Engineering at Purdue University for the Ravi and Eleanor Talwar Rising Star Professorship support. W.W. and P.D.Y. were also supported by NSF under grant No. CMMI-1762698. J.J. and H.W. acknowledge the support from the US National Science Foundation for the microscopy work (DMR-1809520 and ECCS-1902644). A portion of this work was performed at the National High Magnetic Field Laboratory, which is supported by the National Science Foundation Cooperative Agreement No. DMR-1644779 and the State of Florida. C.N. and P.T. acknowledge technical support from National High Magnetic Field Laboratory staff D. Graf, G. Jones, L. Jiao, and A. Suslov.

**Author Contributions**

P.D.Y. conceived the idea and supervised experiments. C.N. fabricated the devices. C.N. and P.T. performed the magneto-transport measurements. Y.W., M.W. and W.W. synthesized the material. G.Q., J.J. and H.W. carried out the TEM/STEM measurements and image analysis. C.N. and G.Q. did the sulfuric acid etching of 2D Te. C.N. and G.Q.

analyzed the data. P.D.Y., C.N. and G.Q. wrote the manuscript and all the authors commented on it.

## Competing Financial Interests

The authors declare no competing financial interests.

## Data Availability

All data needed to evaluate the conclusions in the paper are present in the paper and/or the Supplementary Materials.

## Supplementary Materials

Additional details for nonreciprocal electrical transport in p-type 2D Te, etch pits of left-handed and right-handed 2D Te, nonreciprocal electrical transport in different chirality 2D Te, frequency-dependent nonlinear planar Hall effect, schematics of left- and right-handed Te crystal structures are in the Supplementary Materials.

Supplementary Notes 1-4.

Supplementary Figures S1-S15.

Supporting Video 1-3.

## Corresponding Author

* Peide D. Ye (E-mail: yep@purdue.edu)

# Figures

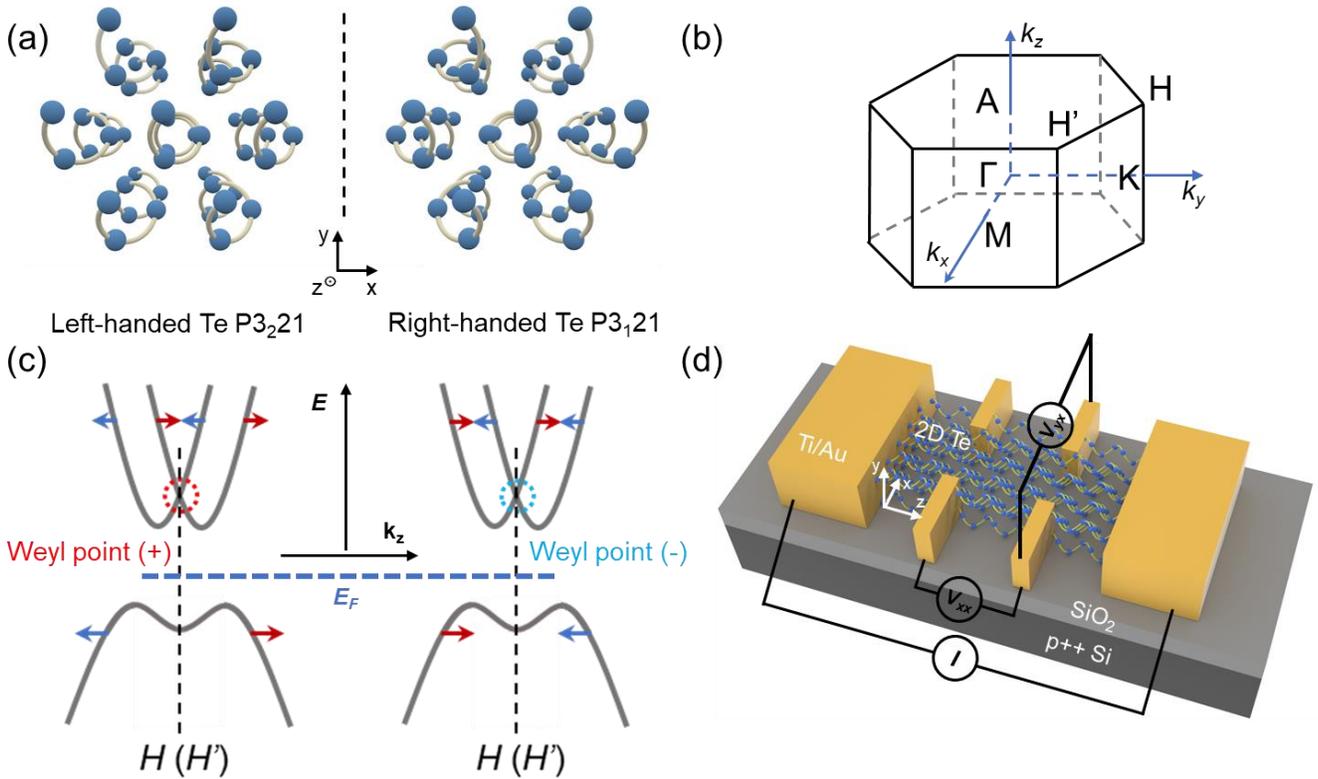

**Figure 1. Chiral crystal and band structure of Te.** (a) Crystal structure of a left-handed Te crystal (P3$_2$21) and a right-handed Te crystal (P3$_1$21). (b) The first Brillouin zone of Te. The conduction band minimum and valence band maximum are located at H and H' point. (c) The electron band structure of chiral Te near the Fermi level. Two enantiomers have the same energy dispersion. However, due to the mirror symmetry, the spin texture and the monopole charge of the Weyl nodes are opposite in left- and right-handed Te. (d) Schematic of the nonlinear electrical transport measurement setup. $I$ is the input ac excitation along the chiral atomic chain direction with a frequency of $\omega$. $V_{zz}$ and $V_{zx}$ are the measured first and second harmonic voltage signals in the longitudinal and transverse directions. The thickness of 2D Te is about 20 nm. The typical length and width of the Hall bar structure is 20 μm and 15 μm, respectively.

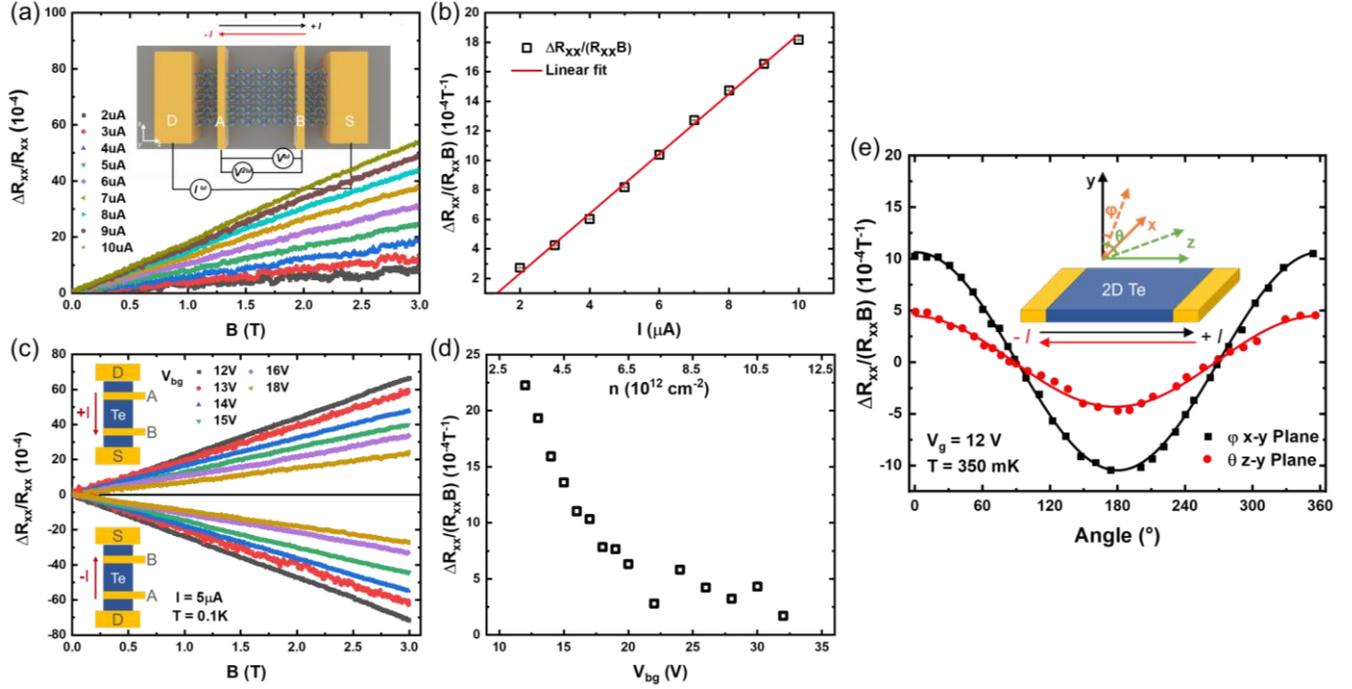

**Figure 2. Nonreciprocal electrical transport in 2D Te.** (a) Magnetic field $B$ dependence of normalized resistance difference $\frac{\Delta R}{R}$ at different current $I$. Inset: schematic of the measurement setup. The magnetic field $B$ is applied in the out-of-plane direction. (b) The current $I$ dependence of $\frac{\Delta R}{RB}$ extracted using the linear fitting of the data from (a) The red line is the linear fitting of $\frac{\Delta R}{RB}$. (c) Magnetic field $B$ dependence of $\frac{\Delta R}{R}$ at different gate voltages (carrier density). Inset: the sketch of the measurement setup. The current was sent from D to S. The voltage was measured between A and B electrodes. (d) The gate voltage dependence of $\frac{\Delta R}{RB}$ extracted using the linear fitting of the data from (c). (e) The angular-dependent nonreciprocal transport in 2D Te in two planes: x-y plane (black) and z-y plane (red). The solid lines are curves fitting by $cos(\varphi)$ and $cos(\theta)$. Inset: schematic of the magnetic field $B$ direction.

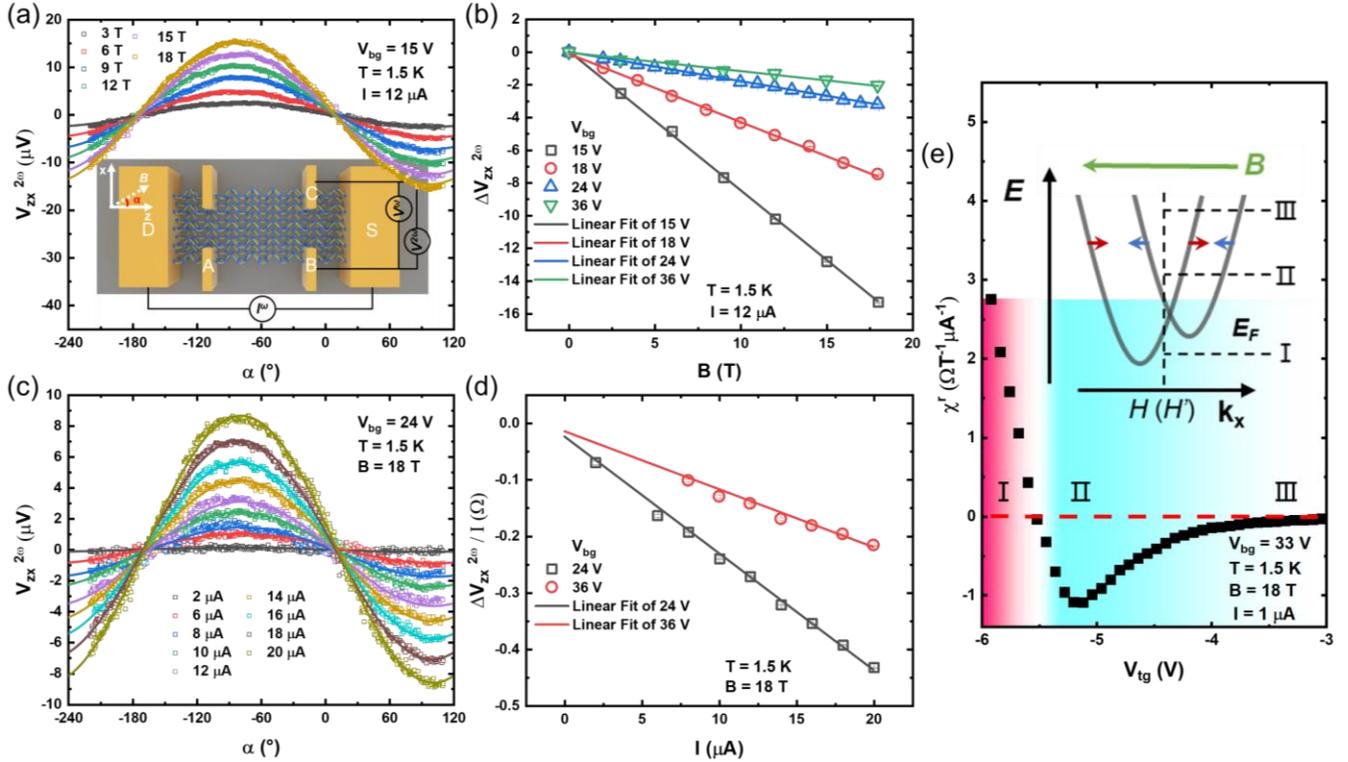

**Figure 3. Nonlinear planar Hall effect in 2D Te.** (a) The second harmonic transverse voltage $V_{zx}^{2\omega}$ as a function of the magnetic field angle α at different magnetic field *B*. The solid lines are curves fitting by $\Delta V_{zx}^{2\omega} \sin(\alpha)$. Inset: schematic of the measurement setup. The magnetic field direction is in-plane. (b) The amplitude of the second harmonic transverse voltage $\Delta V_{zx}^{2\omega}$ at different magnetic field and back gate voltages. The solid lines are linear fitting of $\Delta V_{zx}^{2\omega}$. (c) The second harmonic transverse voltage $V_{zx}^{2\omega}$ as a function of the magnetic field angle α at different current *I*. The solid lines are curves fitting by $\Delta V_{zx}^{2\omega} \sin(\alpha)$. (d) The normalized amplitude of the second harmonic transverse voltage $\Delta V_{zx}^{2\omega}/I$ at different current and back gate voltages. The solid lines are linear fitting of $\Delta V_{zx}^{2\omega}/I$. (e) The top gate voltage dependence of $\chi^r$. Inset: the energy dispersion of right-handed Te conduction band along $k_x$ direction under a magnetic field.

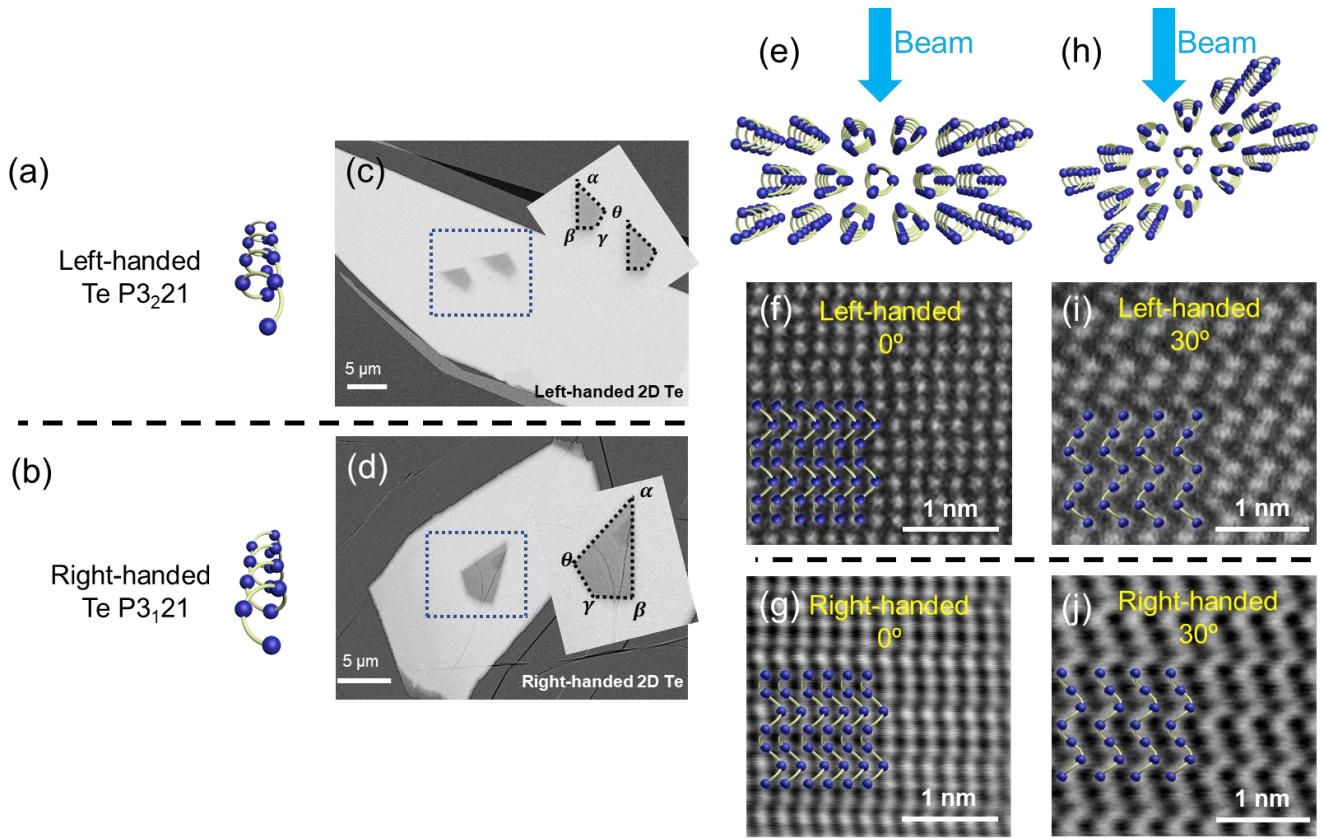

**Figure 4. The chirality of 2D Te.** (a, b) Schematic of the covalently bonded one-dimensional chiral atomic chains with left- and right-handed configuration. (c, d) SEM image of the etch pits in a left-handed 2D Te flake and a right-handed 2D Te flake. (e-j) High resolution scanning transmission electron microscopy (HR-STEM) images of Te flakes with opposite chirality. (e) Configurations of STEM images for $(10\bar{1}0)$ facets. (f, g) Top view STEM images of left-handed and right-handed Te flakes, respectively. Inset: identical top views of both left handed- and right-handed Te atomic model. (h) Configurations of STEM images for $(11\bar{2}0)$ facets. (i, j) STEM images of $(11\bar{2}0)$ facets for left-handed and right-handed Te flakes, respectively. Two images now show distinctive patterns as replicated by rotating both left- and right-handed models along [0001] axis for 30º counterclockwise.

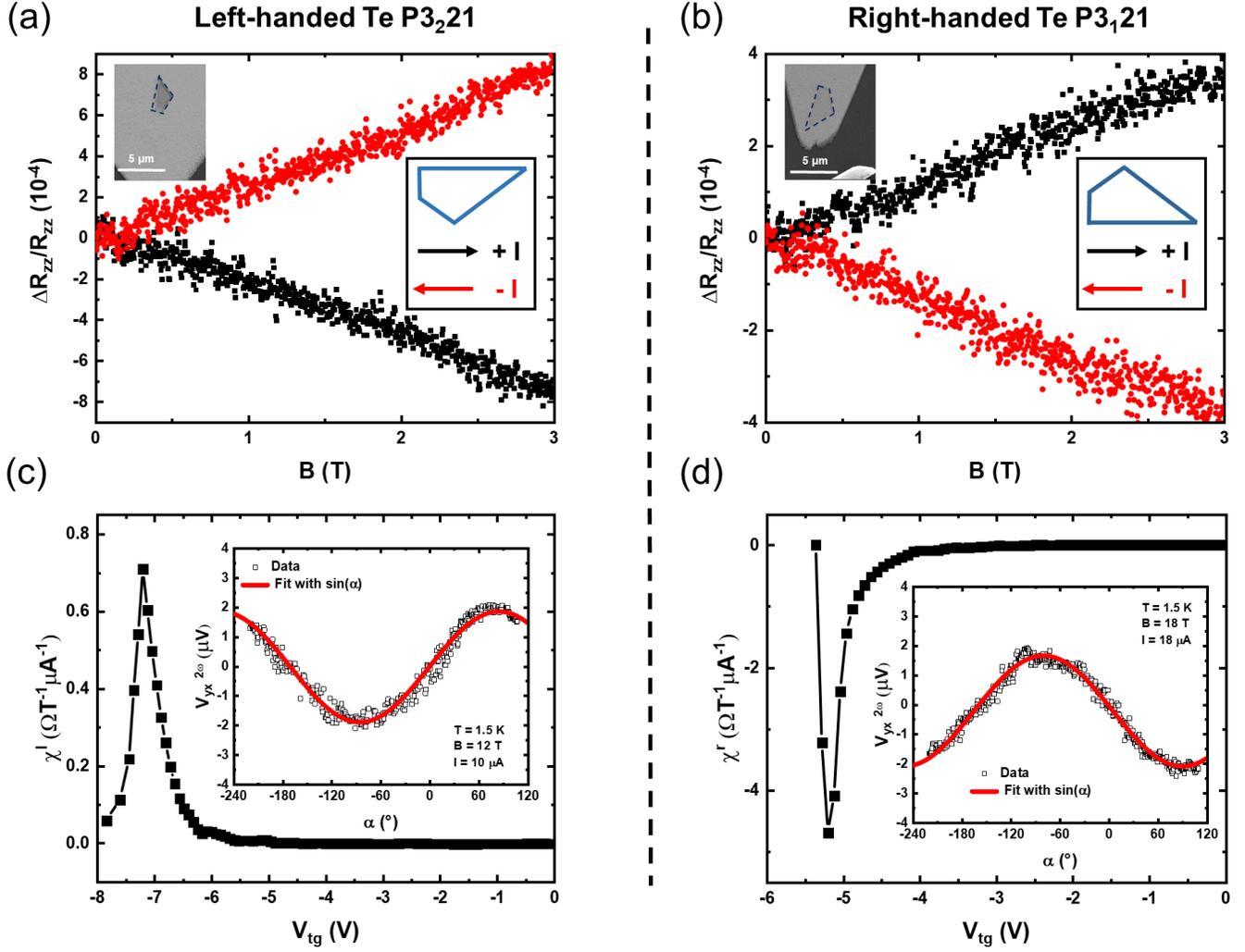

**Figure 5. Chirality-dependent nonlinear electrical transport in 2D Te.** (a, b) The nonreciprocal electrical transport in left- and right-handed 2D Te. Inset: the SEM images of the etch pits and the measurement current direction using the shape of the asymmetry etch pits as reference. (c, d) The nonlinear planar Hall effect in left- and right-handed 2D Te. Inset: the second harmonic transverse voltages as a function of the magnetic field angle $\alpha$. The red lines are curves fitting by $\Delta V_{zx}^{2\omega}\sin(\alpha)$. Both nonreciprocal electrical transport and nonlinear planar Hall effect show opposite signs in 2D Te with different chirality.

**TOC Graphic:**

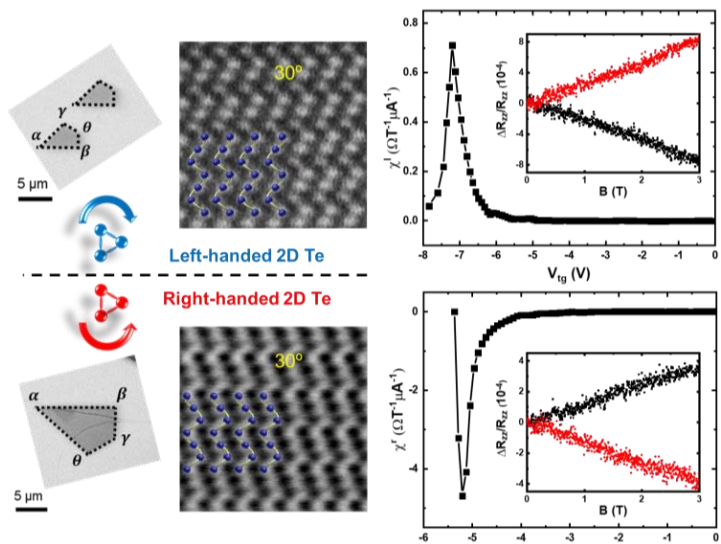

Supporting Information for:

# Tunable Chirality-dependent Nonlinear Electrical Responses in 2D Tellurium

Chang Niu[1,2]†, Gang Qiu[1,2]†, Yixiu Wang[3], Pukun Tan[1,2], Mingyi Wang[3], Jie Jian[4], Haiyan Wang[4], Wenzhuo Wu[3] and Peide D. Ye[1,2]*

[1]*Elmore Family School of Electrical and Computer Engineering, Purdue University, West Lafayette, IN 47907, United States.*

[2]*Birck Nanotechnology Center, Purdue University, West Lafayette, IN 47907, United States.*

[3]*School of Industrial Engineering, Purdue University, West Lafayette, IN 47907, United States.*

[4]*School of Materials Science and Engineering, Purdue University, West Lafayette, Indiana 47907, United States.*

†These authors contributed equally to this work: Chang Niu, Gang Qiu.

*Correspondence and requests for materials should be addressed to P. D. Y. (yep@purdue.edu)

**This file includes:**

**Supplementary Notes:**

1. Phase-sensitive measurement of nonlinear electrical responses.

2. Nonreciprocal electrical transport in p-type 2D Te

3. Etch pits of 2D Te

4. Nonreciprocal electrical transport in Te with different chirality

**Supplementary Figures:**

Figure S1. Spin texture in different materials.

Figure S2. Nonreciprocal electrical transport in p-type 2D Te.

Figure S3. Temperature dependence of the nonreciprocal electrical transport.

Figure S4. Angular-dependent longitudinal nonreciprocal transport with an in-plane magnetic field.

Figure S5. Nonlinear planar Hall effect in positive and negative magnetic field.

Figure S6. Frequency-dependent nonlinear planar Hall effect.

Figure S7. Magnetic field and current dependence of the nonlinear planar Hall effect.

Figure S8. Back gate voltage dependence of the nonlinear planar Hall effect.

Figure S9. Optical images of the etch pits in 2D Te flakes.

Figure S10. SEM pictures of the etch pits in left-handed 2D Te single crystals.

Figure S11. SEM pictures of the etch pits in right-handed 2D Te single crystals.

Figure S12. Origin of the etch pits in 2D Te.

Figure S13. Additional nonreciprocal electrical transport data on left-handed 2D Te.

Figure S14. Additional nonreciprocal electrical transport data on right-handed 2D Te.

Figure S15. Chirality-dependent nonlinear planar Hall effect in 2D Te.

**Other Supplementary Materials for this manuscript include the following:**

**Supplementary videos:**

Supplementary Video 1. Te chains with opposite chirality can potentially have the same top view.

Supplementary Video 2. Front view of Te with opposite chirality under the same transformation of rotating along [0001] axis by 30° counterclockwise.

Supplementary Video 3. Top view of Te with opposite chirality under the same transformation as in Supplementary Video 2. The final top view matches STEM image in **Figure 4i** and **4j**.

**Supplementary Notes:**

**1. Phase-sensitive measurement of nonlinear electrical responses.**

A phase-sensitive measurement is used in the detection of both nonreciprocal electrical transport in the longitudinal direction and the nonlinear planar Hall effect in the transvers direction using the lock-in technique.

In nonreciprocal electrical transport measurement, the ac current with the frequency of $\omega$ can be expressed by $I^\omega = I_0 \sin(\omega t)$. The unidirectional magnetoresistance is described by $R^{l,r}(B, I) = R_0(1 + \beta B^2 + \gamma^{l,r} BI)$. The nonlinear voltage components can be expressed by $V_{xx}^{2\omega}(t) = \gamma^{l,r} R_0 B I_0^2 \sin^2(\omega t) = \frac{1}{2}\gamma^{l,r} R_0 B I_0^2 (1 + \sin(2\omega t - \frac{\pi}{2}))$. In the experiment, the y-component of the second-harmonic voltage drop detected by the lock-in is expressed as $V_{zz}^{2\omega} = \frac{1}{2}\gamma^{l,r} R_0 B I_0^2$ because of the cosine dependence. The ordinary magneto-resistance follows: $V_{zz}^\omega(t) = R_0(1 + \beta B^2) I_0 \sin(\omega t)$. In the experiment, the x-component of the first-harmonic voltage drop detected by the lock-in is expressed as $V_{zz}^\omega = R_0(1 + \beta B^2) I_0$. If $\beta$ is small, according to the definition of the resistance difference between positive and negative current can be written as:

$$\frac{\Delta R}{R} = 2\gamma^{l,r} BI = \frac{4V_{zz}^{2\omega}}{V_{zz}^\omega}$$

where $\Delta R \equiv R(B, I) - R(B, -I)$.

In nonlinear planar Hall effect, the ac current with the frequency of $\omega$ can be expressed by $I^\omega = I_0 \sin(\omega t)$. The nonlinear planar Hall effect is expressed as: $V_{zx}^{2\omega}(t) = R_{zx} I_0^2 \sin^2(\omega t) = \frac{1}{2} R_{zx} I_0^2 (1 + \sin(2\omega t - \frac{\pi}{2}))$. In the experiment, the y-component of the second-harmonic voltage drop detected by the lock-in is expressed as $V_{zx}^{2\omega} = \frac{1}{2} R_{zx} I_0^2$ because of the cosine dependence.

Frequency-dependent of the nonlinear responses are measured shown in **Figure S4a** and **S4b**. The device is measured under same current, magnetic field and gate voltages. The frequency of the current applied is changed from 7 Hz to 87 Hz. No significant change is found, indicating a good ohmic response from the 2D Te device. We exclude the potential effects due to the capacitance in the measurement setup and the device contact, which is sensitive to the current frequency. Furthermore, the x-component and y-component are measured at the same time shown in **Figure S4c**. As discussed previously, the nonlinear planar Hall signal should be cosine dependence ( $V_{zx}^{2\omega}(t) = \frac{1}{2}R_{zx}I_0^2(1 + \sin(2\omega t - \frac{\pi}{2}))$ ), which is the y-component of the measured second-harmonic voltage drop $V_{zx}^{2\omega}$. The y-component has a strong dependence on the magnetic field angle $\alpha$ while the x-component has no dependence on $\alpha$ and is close to zero. Thus, we conclude that the nonlinear response detected by the phase-sensitive lock-in technique originates from the chiral 2D Te crystal.

## 2. Nonreciprocal electrical transport in p-type 2D Te

90 nm $SiO_2$ is used to control the carrier density in p-type 2D Te. The electrical magnetochiral anisotropy is also observed in the valence band of Te (**Figure S1**). Same magnetic field, gate voltage, and current dependence are observed in p-type 2D Te. The amplitude of the nonreciprocal electrical response in n-type 2D Te is about one order of magnitude larger than the amplitude of p-type 2D Te. The enhancement of the nonreciprocal electrical response in n-type Te is related to the Weyl point located at the conduction band edge.

**Figure S2** shows the Temperature dependence of the nonreciprocal electrical transport in p-type 2D from 100 mK to 22 K. The amplitude of nonreciprocal response is decreased with the temperature increasing.

## 3. Etch pits of 2D Te

**Figure S7** shows the optical images of the etch pits in 2D Te flakes. The etch pits are shown in red solid lines. SEM images of the etch pits in left-handed (**Figure S8**) and right-handed (**Figure S9**) 2D Te flakes.

The angles in the etch pits can be calculated using the lattice constants of Te. **Figure S10** shows a simple model for the etch pits in a right-handed Te crystal. The angles can be described as:

$$\tan(\alpha) = \frac{a}{c}, \beta = 90°, \gamma = 90° + \alpha, \theta = 180° - 2\alpha$$

After etching, the shapes of two sides become mirror symmetric (**Figure S10**) which has the identical angles as in etched pits, suggesting that two sides of the flake are also mirror symmetric and hence, possess opposite chirality.

## 4. Nonreciprocal electrical transport in Te with different chirality

The opposite nonreciprocal electrical transport between different chirality is further tested and analyzed in multiple samples. All the devices measured have the same response. Please see **Figure S11** and **S12** for additional electrical measurement.

**Supplementary Figures:**

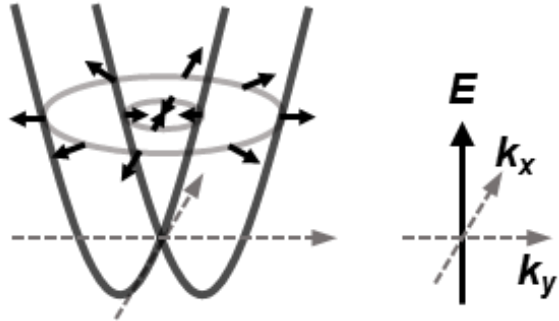

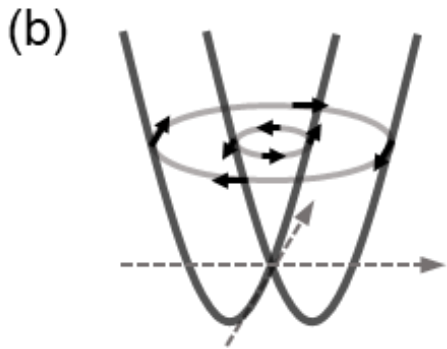
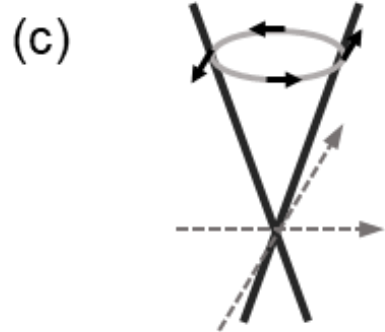

**Figure S1 Spin texture in different materials.** (a) Spin texture of Weyl-type SOC in chiral crystals such as Te conduction band (spin-polarization parallel to the electron momentum direction). (b) Spin texture of Rashba SOC in semiconductors (spin-polarization perpendicular to the electron momentum direction, two Fermi surfaces have opposite spin-polarization). (c) Spin texture in topological insulator surface states (spin-polarization perpendicular to the electron momentum direction).

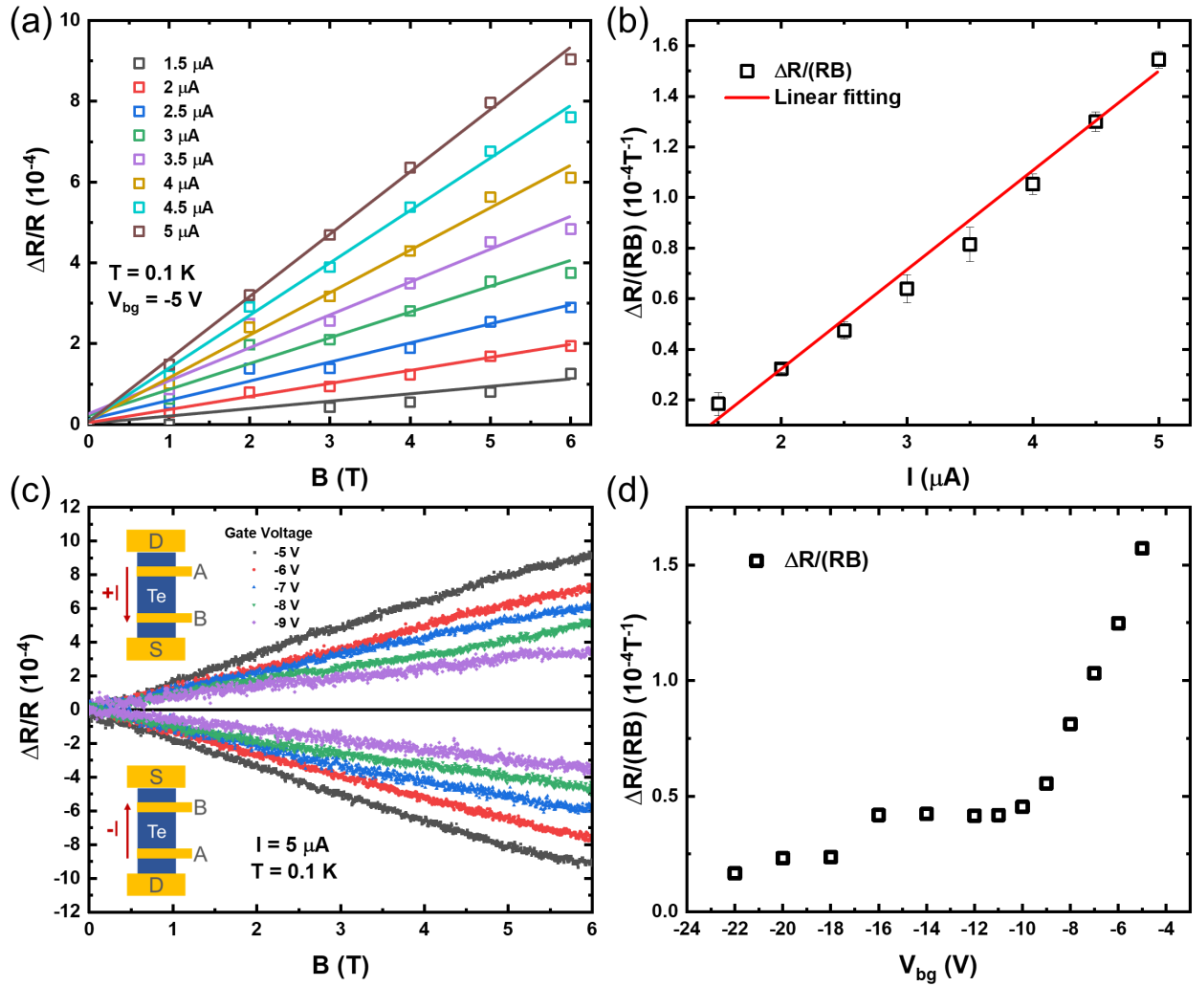

**Figure S2. Nonreciprocal electrical transport in p-type 2D Te.** (a) Magnetic field $B$ dependence of normalized resistance difference $\frac{\Delta R}{R}$ at different current $I$. (b) The current $I$ dependence of $\frac{\Delta R}{RB}$ extracted using the linear fitting of the data from (a) The red line is the linear fitting of $\frac{\Delta R}{RB}$. (c) Magnetic field $B$ dependence of $\frac{\Delta R}{R}$ at different gate voltages (carrier density). Inset: the sketch of the measurement setup. The current was sent from D to S. The voltage was measured between A and B electrodes. (d) The gate voltage dependence of $\frac{\Delta R}{RB}$ extracted using the linear fitting of the data from (c).

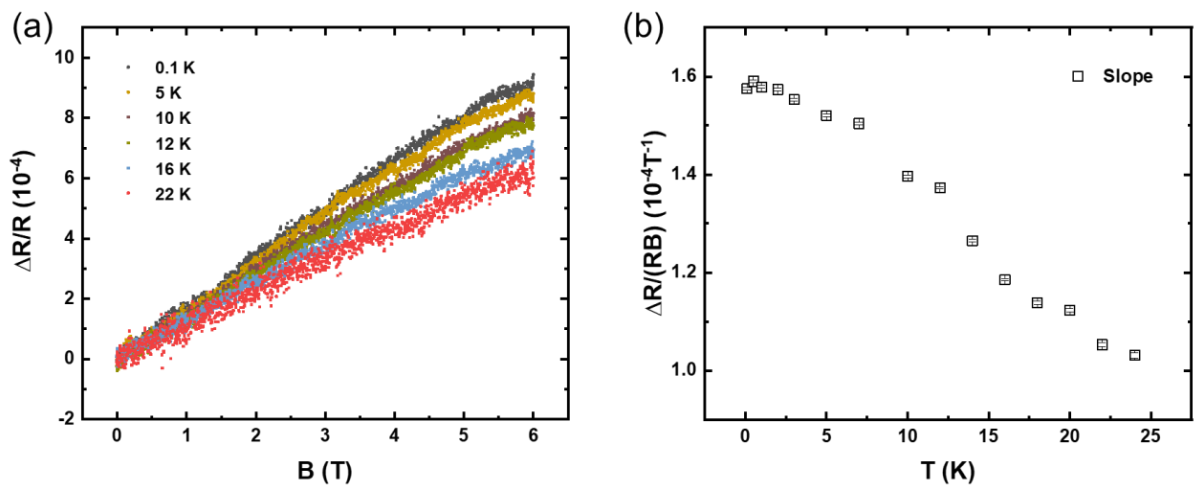

**Figure S3. Temperature dependence of the nonreciprocal electrical transport.** (a) Temperature dependence of the nonreciprocal electrical transport in a p-type 2D Te FET. (b) Temperature dependence of $\frac{\Delta R}{RB}$ extracted using the linear fitting of the data from (a).

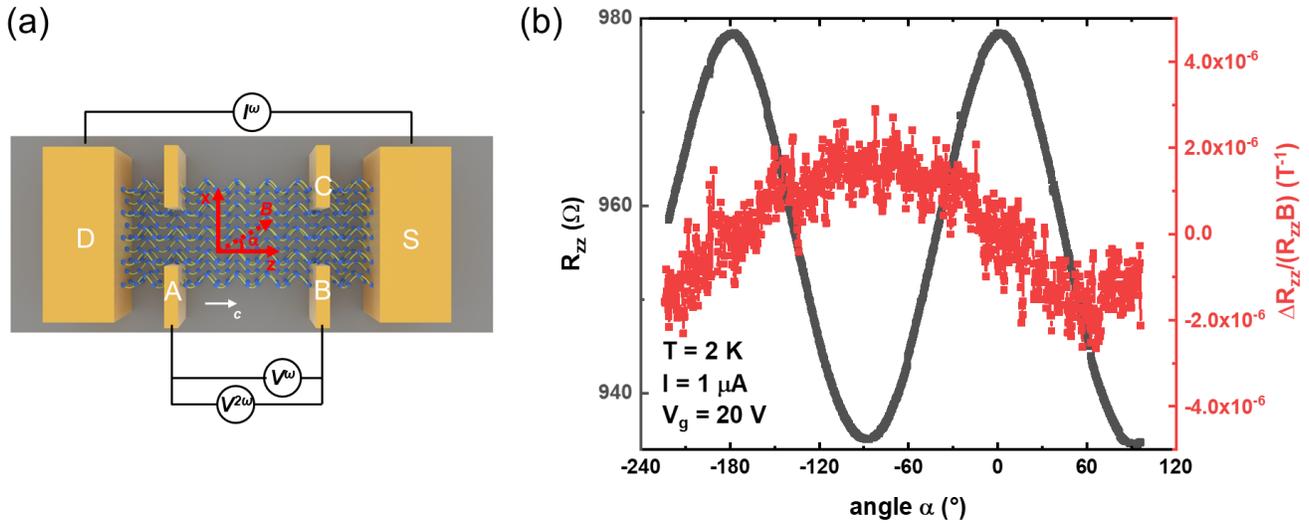

**Figure S4. Angular-dependent longitudinal nonreciprocal transport with an in-plane magnetic field.** (a) Schematic of the in-plane magnetic field measurement setup. (b) First-order (black) and second-order (red) angular-dependent magnetoresistance. Small nonreciprocal current is observed when the magnetic field is perpendicular to the current. No nonreciprocal transport is observed when the magnetic field is parallel to the current.

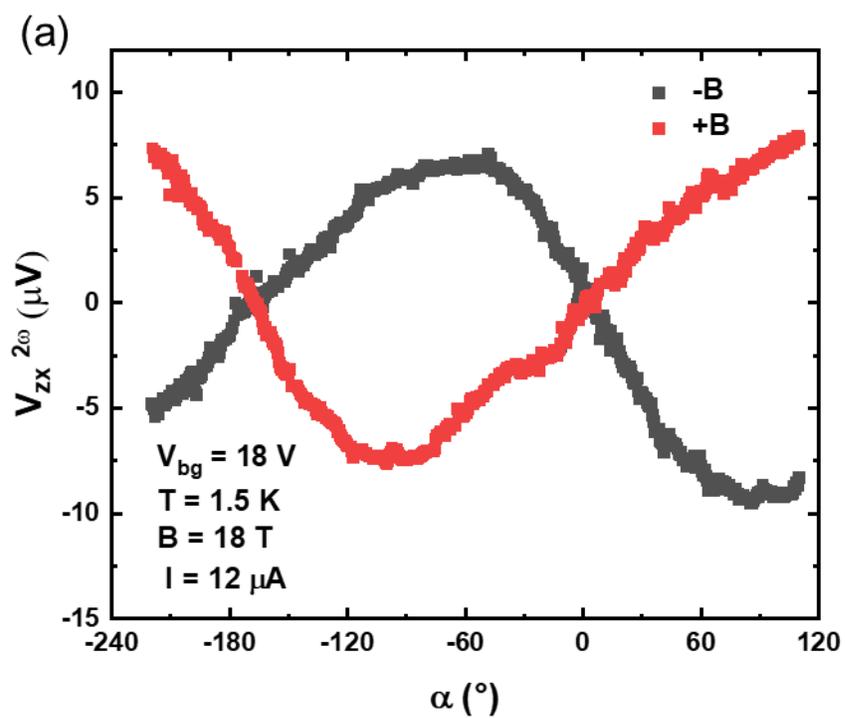

**Figure S5. Nonlinear planar Hall effect in positive and negative magnetic field.** Nonlinear planar Hall signal ($V_{zx}^{2\omega}$) with positive and negative magnetic field.

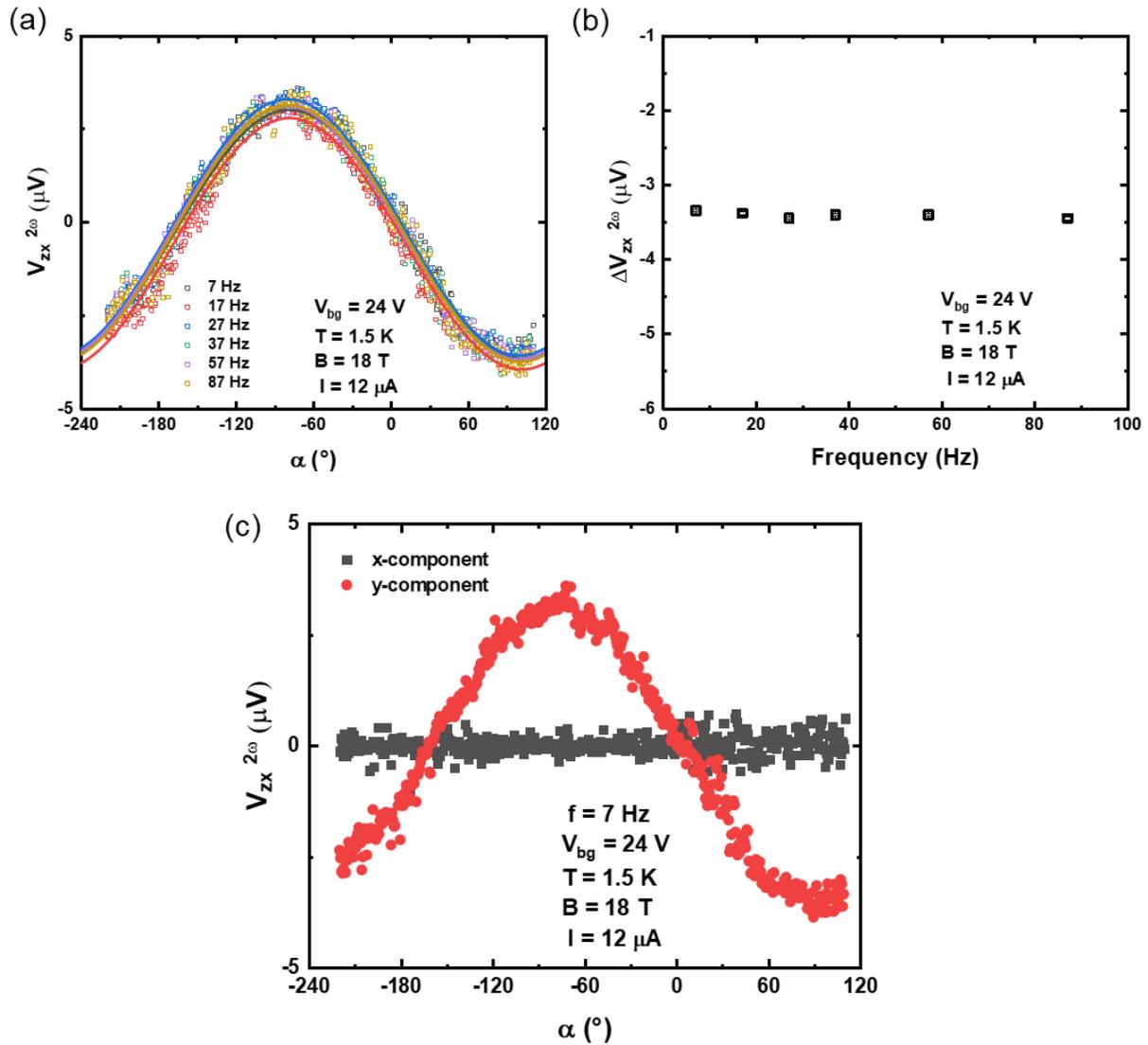

**Figure S6. Frequency-dependent nonlinear planar Hall effect.** (a) Angular-dependent nonlinear planar Hall signal ($V_{zx}^{2\omega}$) at different lock-in frequencies. (b) The amplitude of nonlinear planar Hall effect $\Delta V_{zx}^{2\omega}$ extracted from (a). The signals show no significant frequency dependence. (c) The x-component and y-component of the second-harmonic transverse voltage drop $V_{zx}^{2\omega}$ measured by lock-in at the frequency of 7 Hz.

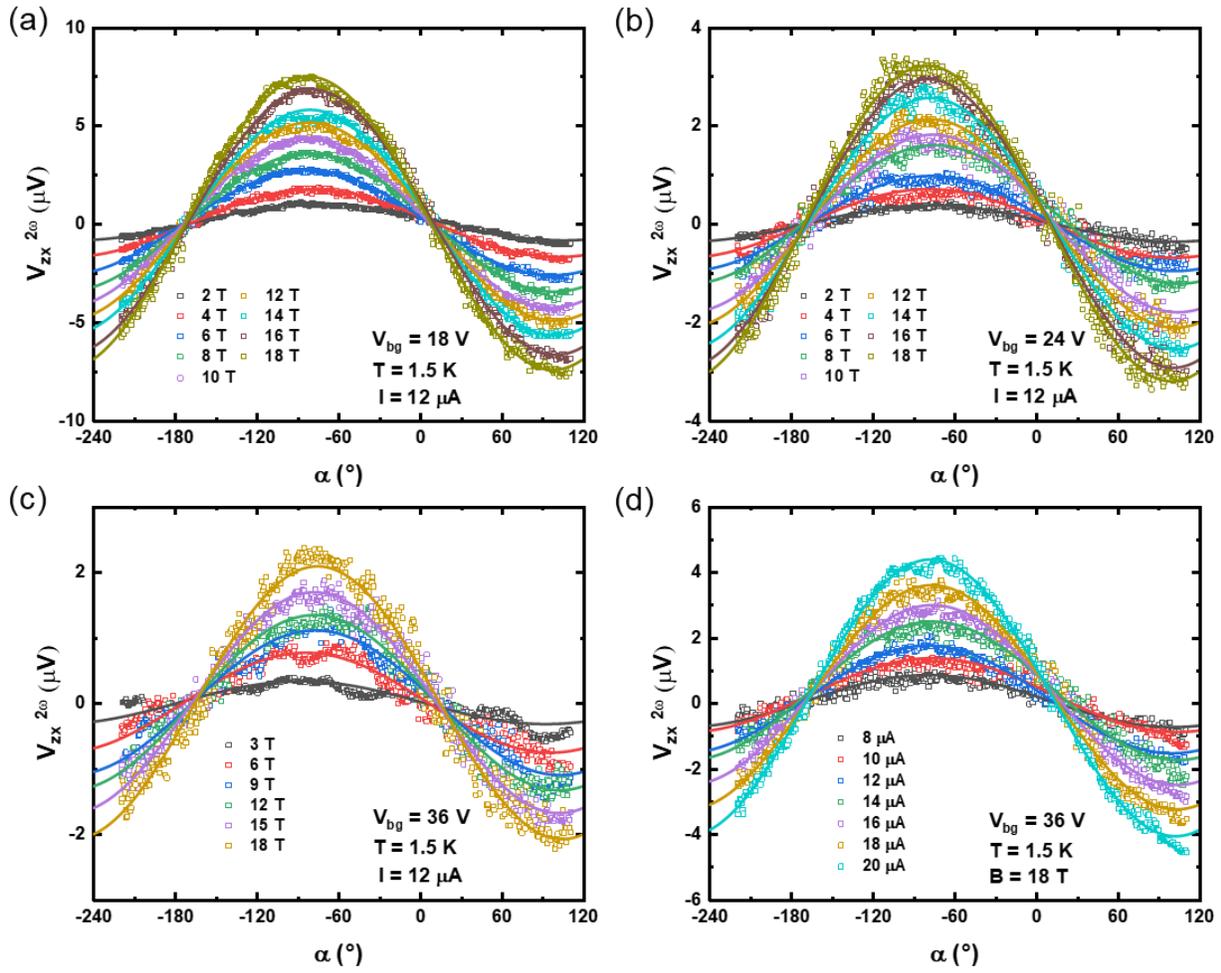

**Figure S7. Magnetic field and current dependence of the nonlinear planar Hall effect.** The second harmonic transverse voltage $V_{zx}^{2\omega}$ as a function of the magnetic field angle $\alpha$ at different magnetic field $B$ (a, $V_{bg} = 18\,V$, b, $V_{bg} = 24\,V$, c, $V_{bg} = 36\,V$) and current $I$ (d, $V_{bg} = 36\,V$). The solid lines are curves fitting by $\Delta V_{zx}^{2\omega} \sin(\alpha)$. The extract nonlinear planar Hall amplitude $\Delta V_{zx}^{2\omega}$ is shown in **Figure 3**.

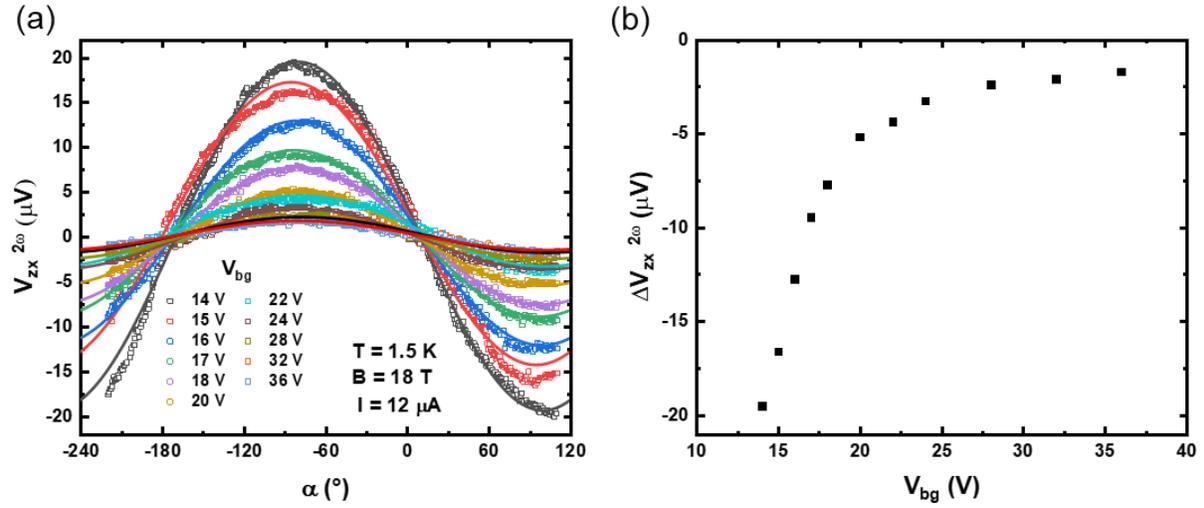

**Figure S8. Back gate voltage dependence of the nonlinear planar Hall effect.** (a) The second harmonic transverse voltage $V_{zx}^{2\omega}$ as a function of the magnetic field angle $\alpha$ at different back gate voltages ($V_{bg}$). The solid lines are curves fitting by $\Delta V_{zx}^{2\omega}\sin(\alpha)$. (b) Back gate voltage $V_{bg}$ dependence of the second harmonic transverse voltage amplitude $\Delta V_{zx}^{2\omega}$ extracted from (a).

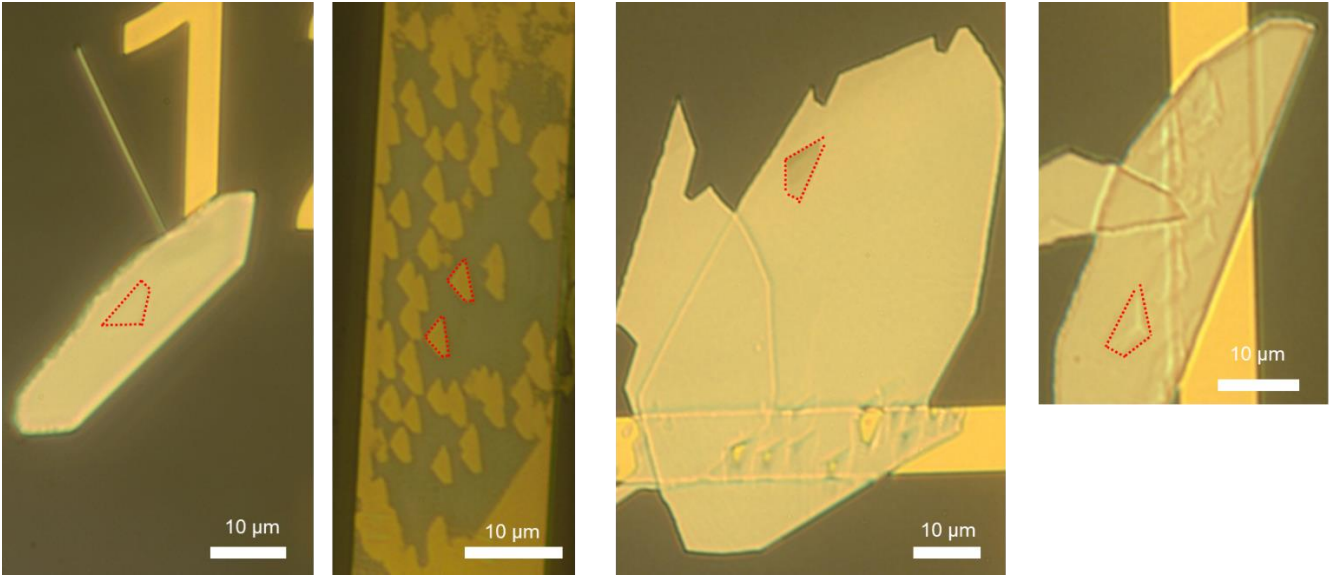

**Figure S9. Optical images of the etch pits in 2D Te flakes.**

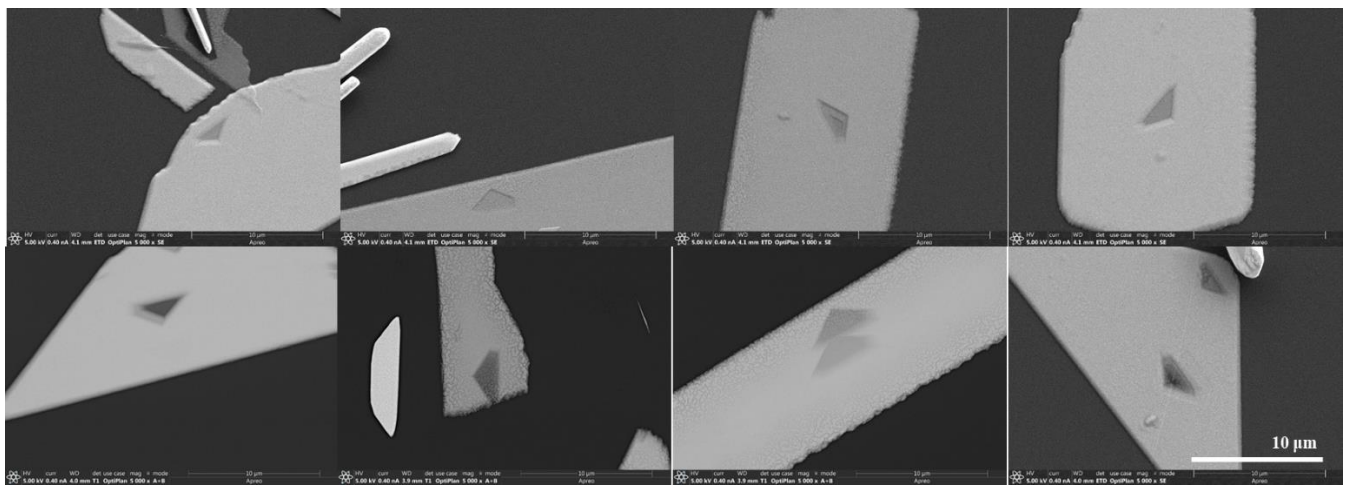

**Figure S10.** SEM pictures of the etch pits in left-handed 2D Te single crystals.

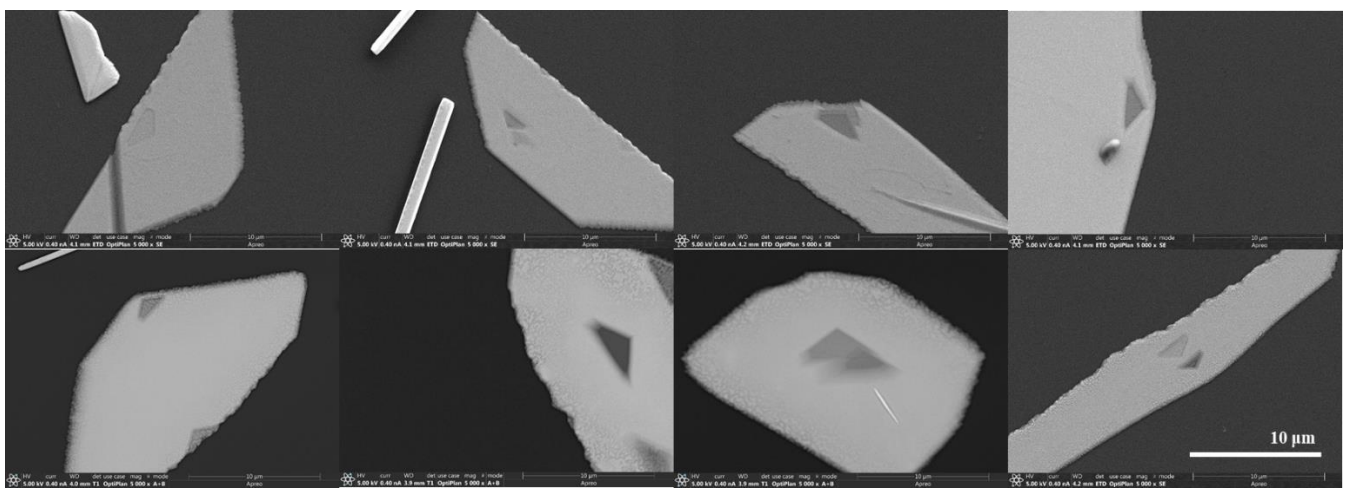

**Figure S11**. SEM pictures of the etch pits in right-handed 2D Te single crystals.

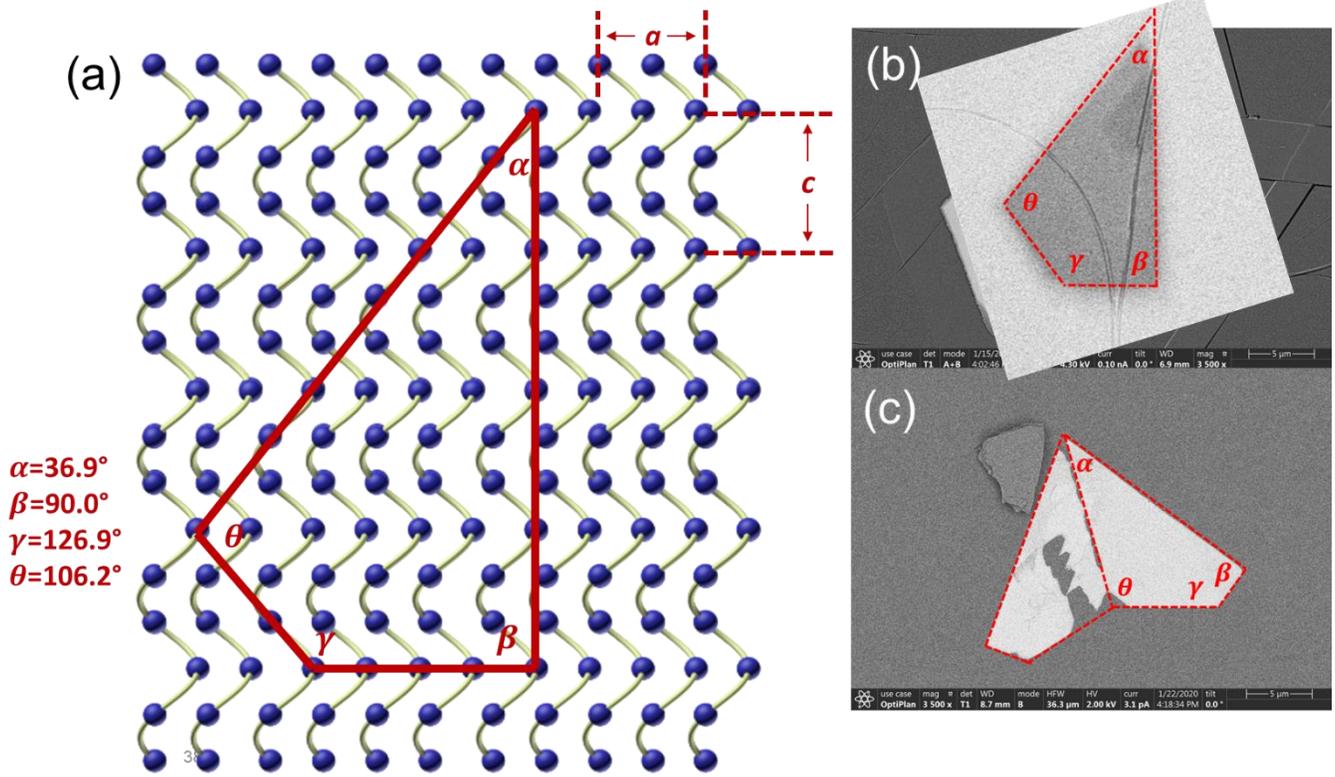

**Figure S12. Origin of the etch pits in 2D Te.** (a) The shape of etch pits (red dash lines) engraved on right-handed Te ($10\bar{1}0$) facet. The etch pits are opposite in a left-handed tellurium flake. (b) Typical etch pits for a right-handed Te. (c) 2D Te flakes after sulfuric etching showing outline shapes similar to etching pits.

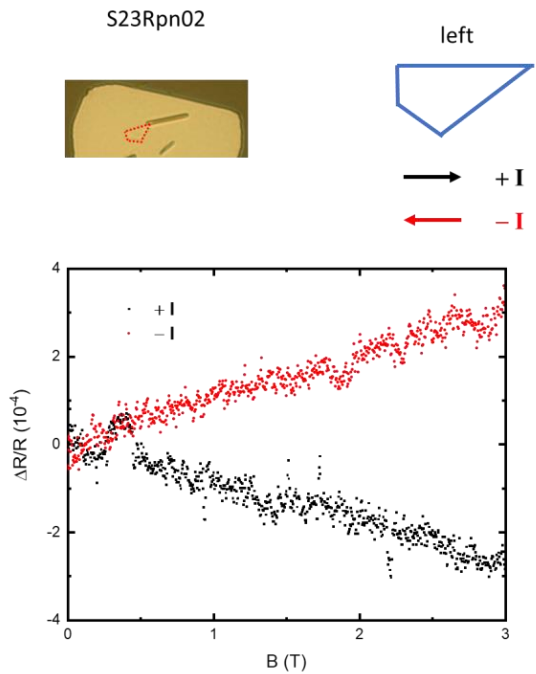

**Figure S13. Additional nonreciprocal electrical transport data on left-handed 2D Te.**

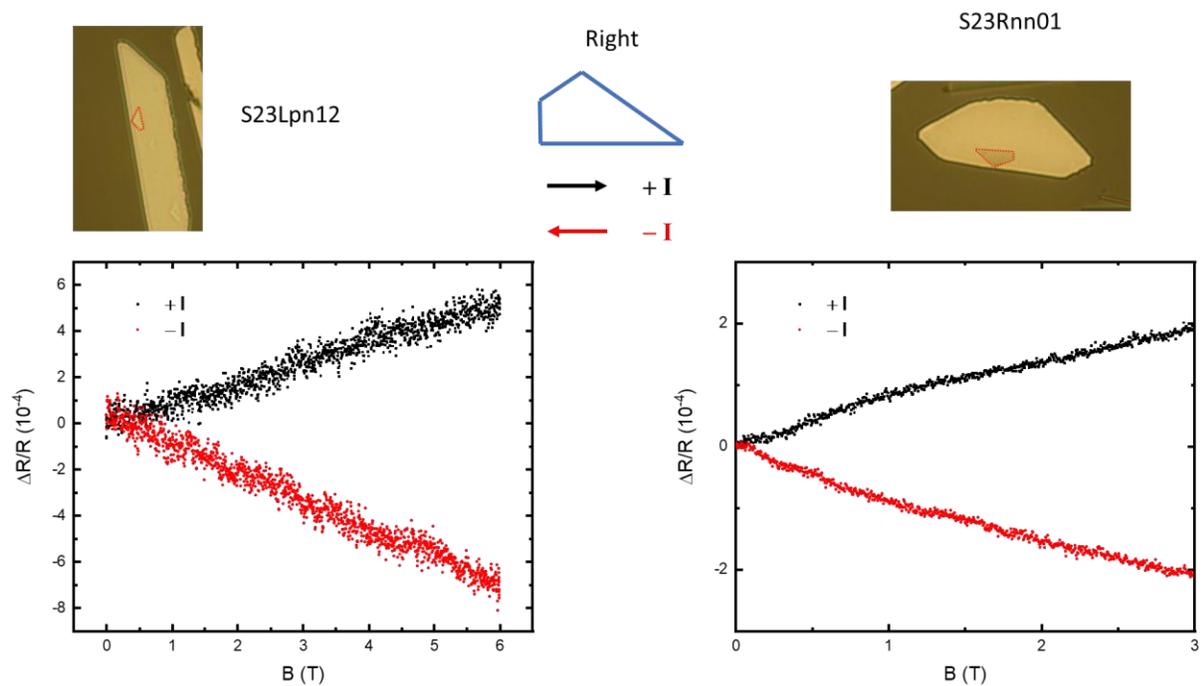

**Figure S14. Additional nonreciprocal electrical transport data on right-handed 2D Te.**

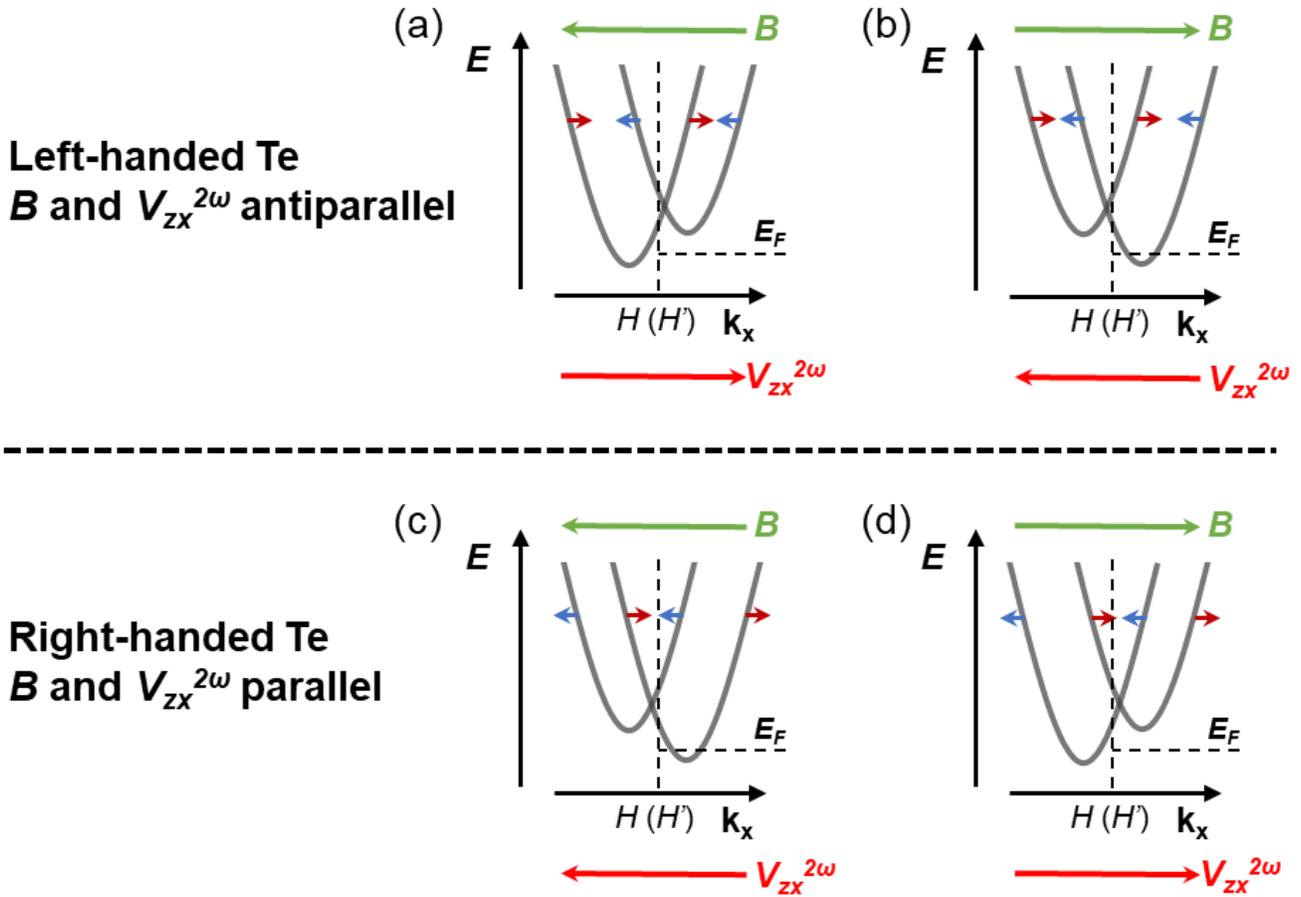

**Figure S15. Chirality-dependent nonlinear planar Hall effect in 2D Te.** The conduction band structure of left- (a, b) and right- (c, d) handed Te along $k_x$ under different magnetic field directions. The red arrows indicate the generated second-order planar Hall current direction. Because of the spin-polarization (spin texture) reverse in left- and right-handed Te, the second-order planar Hall current is parallel (antiparallel) to the magnetic field direction in right- (left-) handed 2D Te.